\documentclass[10pt,conference]{IEEEtran}
\usepackage{amssymb,amsmath}
\usepackage{graphics}
\usepackage{epsfig}
\usepackage{graphicx}

\newtheorem{theorem}{Theorem}
\newtheorem{lemma}{Lemma}

\begin{document}

\title{Delay-Throughput Tradeoff for Supportive Two-Tier Networks}
\author{\authorblockN{Long Gao\authorrefmark{1}, Rui Zhang\authorrefmark{2}, Changchuan Yin\authorrefmark{1}, Shuguang Cui\authorrefmark{1}}
\authorblockA{\authorrefmark{1}Department of Electrical and Computer Engineering\\ Texas A\&M University\\ College Station, TX, 77843\\ \{lgao, ccyin, cui\}@ece.tamu.edu}
\authorblockA{\authorrefmark{2}Institute for Infocomm Research, A*STAR, Singapore\\rzhang@i2r-star.edu.sg}}

\maketitle

\begin{abstract}
Consider a static wireless network that has two tiers with
different priorities: a primary tier vs. a secondary tier. The
primary tier consists of randomly distributed legacy nodes of
density $n$, which have an absolute priority to access the
spectrum. The secondary tier consists of randomly distributed
cognitive nodes of density $m=n^\beta$ with $\beta\geq 2$, which
can only access the spectrum opportunistically to limit the
interference to the primary tier. By allowing the secondary tier
to route the packets for the primary tier, we show that the
primary tier can achieve a throughput scaling of
$\lambda_p(n)=\Theta\left(1/\log n\right)$ per node and a
delay-throughput tradeoff of $D_p(n)=\Theta\left(\sqrt{n^\beta\log
n}\lambda_p(n)\right)$ for $\lambda_p(n)=O\left(1/\log n\right)$,
while the secondary tier still achieves the same optimal
delay-throughput tradeoff as a stand-alone network.
\end{abstract}

\IEEEpeerreviewmaketitle

\section{Introduction}

The explosive growth of large-scale wireless applications
motivates people to study the fundamental limits over wireless
networks. Consider a randomly distributed wireless system with
density $n$ over a unit area, where the nodes are randomly grouped
into one-to-one source-destination (S-D) pairs. Initiated by the
seminal work in~\cite{Gupta:Capacity}, the throughput scaling laws
for such a network have been studied extensively in the
literature~\cite{Francheschetti:Closing}-\cite{Matthias:Mobility}.
For static networks, it is shown~\cite{Gupta:Capacity} that the
traditional multi-hop transmission strategy can achieve a
throughput scaling of $\Theta\left(1/\sqrt{n\log
n}\right)$\footnote{We use the following notations throughout this
paper: i) $f(n)=O(g(n))$ means that there exists a constant $c$
and integer $N$ such that $f(n)<cg(n)$ for $n>N$; ii)
$f(n)=\Omega(g(n))$ means that $g(n)=O(f(n))$; iii)
$f(n)=\Theta(g(n))$ means that $f(n)=O(g(n))$ and $g(n)=O(f(n))$;
iv) $f(n)=o(g(n))$ means $f(n)/g(n)\rightarrow 0$ as
$n\rightarrow\infty$.} per S-D pair. Besides the throughput, the
packet delay is another key performance metric in wireless
networks. The delay-throughput tradeoffs for static and mobile
networks have been investigated
in~\cite{Michael:Capacity}-\cite{Gamal:Optimal}. Specifically, for
static networks, it is shown in~\cite{Gamal:Optimal} that the
optimal delay-throughput tradeoff is given by
$D(n)=\Theta(n\lambda(n))$ for $\lambda(n)=O\left(1/\sqrt{n\log
n}\right)$, where $D(n)$ and $\lambda(n)$ are the delay and
throughput per S-D pair, respectively.

The aforementioned literature focuses on the delay and throughput
scaling laws for a single network. Recently, the emergence of
cognitive radio networks leads to the necessity of extending the
result from a single network to multiple overlaid networks.
Consider a licensed primary network and a cognitive secondary
network coexisting in a unit area. The primary network has the
absolute priority to use the spectrum, while the secondary network
can only access the spectrum opportunistically such that the
resulted interference to the primary network is tolerable. Based
on such assumptions, the delay and throughput performance for the
two overlaid networks have been studied
in~\cite{Jeon:Cognitive}~\cite{Yin:Scaling}.
In~\cite{Jeon:Cognitive}, it has been shown that by defining a
preservation region around each primary node, both networks can
achieve the same throughput scaling law given
in~\cite{Gupta:Capacity} as a stand-alone wireless network.
In~\cite{Yin:Scaling}, it has been further shown that both
networks achieve the same delay-throughput tradeoff as the optimal
one established in~\cite{Gamal:Optimal}. However, these existing
results are obtained without considering possible positive
interactions between the primary network and the secondary
network. In practice, the secondary network, which is usually
deployed after the existence of the primary network for
opportunistic spectrum access, can transport not only its own data
packets but also the packets for the primary network. As such, a
natural question arises whether the throughput and/or delay
performance of the primary network can be improved via the aid of
the newly-added secondary network, without the need of
modifications over the existing primary protocol. Meanwhile, we
study whether the the secondary network is still capable of
keeping the same throughput and delay scaling laws as in a
stand-alone network.

In particular, we consider the primary network and the secondary
network as two coexisting tiers with different priorities (a
legacy tier vs. a cognitive tier) in a wireless system, where each
of them has its own data packets to transmit such that we need to
evaluate the performance of the two tiers respectively. In such a
system, we allow the secondary tier to supportively relay the data
packets for the primary tier in an opportunistic way, while the
primary tier is only required to transmit its own data.
Furthermore, we require the primary protocol to keep the same as
in a legacy single-tier network given the fact that the primary
tier may already exist before the secondary tier is added in.
Based on these assumptions, we deploy a similar primary protocol
to those
in~\cite{Gupta:Capacity}~\cite{Jeon:Cognitive}~\cite{Yin:Scaling}
and propose a new multi-hop transmission protocol for the
secondary tier. We show that when the secondary tier has a higher
density than the primary tier, i.e., $m=n^\beta$ with $\beta\geq
2$, the throughput scaling law of the primary tier can be
significantly improved from $\Theta\left(1/\sqrt{n \log n}\right)$
to $\Theta\left(1/\log n\right)$ per S-D pair with the aid of the
secondary tier, while the secondary tier can still achieve the
throughput scaling law of $\Theta\left(1/\sqrt{m \log m}\right)$
per S-D pair as a stand-alone network. Furthermore, we investigate
the delay performance and the delay-throughput tradeoff within
each tier.

The rest of the paper is organized as follows. The system model is
described in Section II. The proposed protocols for the primary
and secondary tiers are described in Section III. The delay and
throughput scaling laws for the secondary tier are shown in
Section IV. The delay and throughput scaling laws for the primary
tier are shown in Section V. Finally, Section VI summarizes our
conclusions.

\section{System Model}

In this section, we first describe the network model and the
system assumptions, and then define the transmission rate and
throughput. In the following, we use $p(E)$ to represent the
probability of event $E$, and claim that an event $E_n$ occurs
with high probability (w.h.p.) if $p(E_n)\to 1$ as $n\to \infty$.
\vspace{-1 mm}
\subsection{Network Model}

Consider two network tiers over a unit square. The nodes of the
primary tier, so-called primary nodes, are distributed according
to a Poisson point process (PPP) of density $n$ and randomly
grouped into one-to-one source-destination (S-D) pairs. Likewise,
the nodes of the secondary tier, so-called secondary nodes, are
distributed according to a PPP of density $m$ and randomly
grouped into S-D pairs. We assume that the density of the
secondary tier is higher than that of the primary tier, i.e.,
\begin{equation} \label{density}
m=n^\beta
\end{equation}
where we consider the case $\beta\geq 2$. The primary tier and the
secondary tier share the same time, frequency, and space, but have
different priorities to access the spectrum: The former one is the
licensed user of the spectrum and thus has a higher priority; and
the latter one can only opportunistically access the spectrum to
limit the resulting interference to the primary tier.

For the wireless channel, we only consider the large-scale
pathloss and ignore the effects of shadowing and small-scale
multipath fading. As such, the channel power gain $g(r)$ is given
as
\begin{equation} \label{pathloss}
g(r)=r^{-\alpha}
\end{equation}
where $r$ is the distance between the transmitter (TX) and the
corresponding receiver (RX), and $\alpha >2$ denotes the pathloss
exponent.
\vspace{-1 mm}
\subsection{Interaction Model}

As shown in the previous
work~\cite{Jeon:Cognitive}~\cite{Yin:Scaling}, although the
opportunistic data transmission in the secondary network does not
degrade the scaling law of the primary network, it may reduce the
throughput in the primary tier by a constant factor due to the
fact that the interference from the secondary network to the
primary network cannot be reduced to zero. To completely
compensate the throughput degradation or even improve the
throughput scaling law of the primary tier in a two-tier setup, we
could allow certain positive interactions between the two tiers.
Specifically, we assume that the secondary tier is willing to
route packets for the primary tier, while the primary tier is not
assumed to do so. In particular, when a primary node sends
packets, the surrounding secondary nodes could pretend to be
primary nodes to relay the packets. After a secondary node
receives the packets from a primary node, it may chop the packets
into smaller pieces suitable for secondary-tier transmissions. The
small data pieces will be assembled before they are delivered to
the primary destination nodes. Note that, these ``fake'' primary
nodes do not have the same priority as the real primary nodes in
terms of spectrum access, i.e., they can only use the spectrum
opportunistically in the same way as a regular secondary node. As
such, the primary tier is expected to achieve a more efficient
throughput scaling law. The assumption of allowing packet
exchanges between the two tiers is the essential difference from
the models in~\cite{Jeon:Cognitive}~\cite{Yin:Scaling}.
\vspace{-1 mm}
\subsection{Throughput and Delay}

The \emph{throughput per S-D pair} is defined as the average data
rate that each source node can transmit to its chosen destination
as in~\cite{Jeon:Cognitive}~\cite{Yin:Scaling}, which is a
function of the network density. Besides, the \emph{sum
throughput} is defined as the product between the throughput per
S-D pair and the number of S-D pairs in the network. In the
following, we use $\lambda_{p}(n)$ and $\lambda_{s}(m)$ to denote
the throughput per S-D pair for the primary tier and the secondary
tier, respectively; we use $T_{p}(n)$ and $T_{s}(m)$ to denote the
sum throughputs for the primary tier and the secondary tier,
respectively.

The delay of a primary packet is defined as the average number of
primary time slots that it takes to reach the primary destination
node after the departure from the primary source node. Similarly,
we define the delay of a secondary packet as the average number of
secondary time slots for the packet to travel from the secondary
source node to the secondary destination node. We use $D_p(n)$ and
$D_s(m)$ to denote packet delays for the primary tier and the
secondary tier, respectively. For simplicity, we use a fluid
model~\cite{Gamal:Optimal} for the delay analysis, in which we
divide each time slot to multiple packet slots and the size of the
data packets can be scaled down to be arbitrarily small with the
increase of node density.

\section{Network Protocols}

In this section, we describe the proposed protocols for the
primary tier and the secondary tier, respectively. The primary
tier deploys the same time-slotted multi-hop transmission scheme
as those for the primary network
in~\cite{Jeon:Cognitive}~\cite{Yin:Scaling}, while the secondary
tier adapts its protocol to the primary transmission scheme. We
start with the protocol for the primary tier, and then describe
the protocol for the secondary tier.
\vspace{-1 mm}
\subsection{The Primary Protocol}
The primary protocol follows the multi-hop transmission scheme
described in~\cite{Jeon:Cognitive}~\cite{Yin:Scaling}, which is
similar to the one proposed in~\cite{Gupta:Capacity}. The main
sketch of the protocol is given as follows:
\begin{itemize}
\item Divide the unit square into small-square primary cells with
size $a_p(n)$. In order to maintain the full connectivity within
the primary tier even without the aid of the secondary tier, we
have $a_p(n)\geq 2\log n/n$ such that each cell has at least one
primary node w.h.p..

\item Group every 64 primary cells into a primary cluster. The
cells in each primary cluster take turns to be active in a
round-robin fashion. We divide the transmission time into TDMA
frames, where each frame has 64 time slots that correspond to the
number of cells in each primary cluster. Note that the number of
primary cells in a primary cluster has to be no less than 64 such
that we can appropriately arrange the preservation regions and the
collection regions, which will be formally defined in the next
section for the secondary protocol.

\item Define the data path along which the packets are routed from
the source node to the destination node: The data path follows a
horizontal line and a vertical line connecting the source node and
the destination node, which is the same as that defined
in~\cite{Jeon:Cognitive}~\cite{Yin:Scaling}. Pick an arbitrary
node within a primary cell as the designated relay node, which is
responsible for relaying the packets of all the data paths passing
through the cell.

\item When a primary cell is active, each primary source node in
it takes turns to transmit one of its own packets. Afterwards, the
designated relay node transmits one packet for each of the S-D
paths passing through the cell. The above packet transmissions
follow a TDMA pattern within the designated primary time slot. For
each packet, if the destination node is found in the adjacent
cell, the packet will be directly delivered to the destination.
Otherwise, the packet is forwarded to the designated relay node in
the adjacent cell along the data path.

\item At each transmission, the TX node can only transmit to a
node in its adjacent cells with power of
$Pa_{p}^{\frac{\alpha}{2}}(n)$, where $P$ is a constant.
\end{itemize}

Note that the protocol for the primary tier does not need to
change no matter the secondary tier is present or not. When the
secondary tier is absent, the primary tier can achieve the
throughput scaling law given in~\cite{Gupta:Capacity} and the
optimal delay-throughput tradeoff given in~\cite{Gamal:Optimal}.
When the secondary tier is present as shown in Section V, the
primary tier can achieve a better throughput scaling law and a
different delay-throughput tradeoff with the aid of the secondary
tier.
\vspace{-1 mm}
\subsection{The Secondary Protocol}

\begin{figure}
\centering
\includegraphics[width=1.9in]{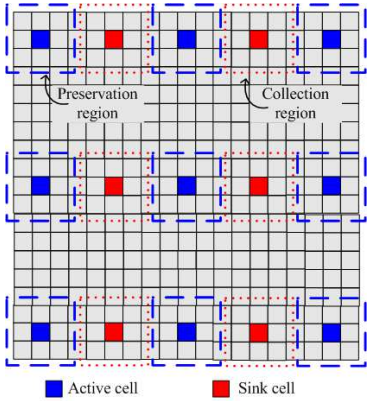}
\caption{Preservation regions and collection regions.}
\label{preservation} \vspace{-15pt}
\end{figure}

Next we describe the protocol for the secondary tier. We assume
that the secondary nodes have the necessary cognitive features to
``pretend'' as primary nodes such that they could be chosen as the
designated primary relay nodes within a particular primary cell.
Once a secondary node is chosen to be a designated primary relay
node for primary packets, it keeps silent during active primary
time slots such that only primary source nodes transmit their
packets at a given primary time slot. Instead, it only relays
primary packets opportunistically by sharing the time resource
with other secondary transmissions. Furthermore, we use the
time-sharing technique to guarantee successful packet deliveries
from the secondary nodes to the primary destination nodes as
follows. We divide each secondary frame into three equal-length
subframes, such that each of them has the same length as one
primary time slot. The first subframe is used to transmit the
secondary packets within the secondary tier. The second subframe
is used to relay the primary packets to the next relay nodes.
Accordingly, the third subframe of each secondary frame is used to
deliver the primary packets from the intermediate destination
nodes\footnote{An ``intermediate'' destination node of a primary
packet within the secondary tier is a chosen secondary node in the
neighboring cells of a particular primary cell within which the
final primary destination node is located.} in the secondary tier
to their final destination nodes in the primary tier.
Specifically, for the first subframe, we use the following
protocol:
\begin{itemize}
\item Divide the unit area into square secondary cells with size
$a_s(m)$. In order to maintain the full connectivity within the
secondary tier, we have to guarantee $a_s(m) \geq 2\log m/m$ with
a similar argument to that in the primary tier.

\item Group the secondary cells into secondary clusters, with each
secondary cluster of 64 cells. Each secondary cluster also follows
a 64-TDMA pattern to communicate. That is, The first subframe is
divided into 64 secondary time slots.

\item Define a preservation region as nine primary cells centered
at an active primary TX and a layer of secondary cells around
them, shown as the square with dashed edges in
Fig.~\ref{preservation}. Only the secondary TXs in an active
secondary cell outside all the preservation regions can transmit
data packets; otherwise, they buffer the packets until the
particular preservation region is cleared. When an active
secondary cell is outside the preservation regions in the first
subframe, it allows the transmission of one packet for each
secondary source node and each S-D path passing through the cell
in a time-slotted pattern within the active secondary time slot w.h.p..

\item At each transmission, the active secondary TX node can only
transmit to a node in its adjacent cells with power of $P
a_{s}^{\frac{\alpha}{2}}(m)$.
\end{itemize}

In the second subframe, only secondary nodes who carry primary
packets take the time resource to transmit. Note that each primary
packet is broadcasted from the primary source node to its
neighboring primary cells where we assume that there are $N$
secondary nodes in the neighboring cell along the data path
successfully decode the packet and ready to relay. In particular,
each secondary node relays $1/N$ portion of the primary packet to
the intermediate destination node in a multi-hop fashion, and the
value of $N$ is set as
\begin{equation}\label{relaynumber}
    N=\Theta\left(\sqrt{\frac{m}{\log m}}\right).
\end{equation}
From \emph{Lemma 1} in Section V, we can guarantee that there are
more than $N$ secondary nodes in each neighboring primary cell of
a primary TX w.h.p. when $\beta\geq 2$. The specific transmission
scheme in the second subframe is the same as that in the first
subframe, where the secondary subframe is all divided into 64 time
slots and all the traffic is for primary packets.

At the intermediate destination nodes, the received primary packet
segments are assembled into the original primary packets. Then in
the third subframe, we use the following protocol to deliver the
packets to the primary destination nodes:
\begin{itemize}
\item Define a collection region as nine primary cells and a layer
of secondary cells around them, shown as the square with dotted
edges in Fig.~\ref{preservation}, where the collection region is
located between two preservation regions along the horizontal line
and they are not overlapped with each other.

\item Deliver the primary packets from the intermediate
destination nodes in the secondary tier to the corresponding
primary destination nodes in the sink cell, which is defined as
the center primary cell of the collection region. The primary
destination nodes in the sink cell take turns to receive data by
following a TDMA pattern, where the corresponding intermediate
destination node in the collection region transmits by pretending
as a primary TX node. Given that the third subframe is of an equal
length to one primary slot, each primary destination node in the
sink cell can receive one primary packet from the corresponding
intermediate destination node.

\item At each transmission, the intermediate destination node
transmits with the same power as that for a primary node, i.e., $P
a_{p}^{\frac{\alpha}{2}}(n)$.
\end{itemize}

\section{Delay and Throughput Analysis for the Secondary Tier}

In this section, we discuss the delay and throughput scaling laws
for the secondary tier. According to the protocol for the
secondary tier, we split the time frame into three equal-length
fractions and use one of them for the secondary packet
transmissions. Since the above time-sharing strategy only incurs a
constant penalty (i.e., 1/3) on the achievable throughput and
delay within the secondary tier, the throughput and delay scaling
laws are the same as those given in~\cite{Yin:Scaling}, which are
summarized by the following theorems.
\begin{theorem}
With the secondary protocol defined in Section III, the secondary
tier can achieve the following throughput per S-D pair and sum
throughput w.h.p.:
\begin{equation}\label{sthroughput1}
    \lambda_s(m)=\Theta\left(\frac{1}{m\sqrt{a_s(m)}}\right)
\end{equation}
and
\begin{equation}\label{sthroughput2}
    T_s(m)=\Theta\left(\frac{1}{\sqrt{a_s(m)}}\right),
\end{equation}
where $a_s(m)\geq 2\log m/m$ and the specific value of $a_s(m)$ is
determined by $a_p(n)$ as shown later in~(\ref{cellsize}).
\end{theorem}
\begin{theorem}
With the secondary protocol defined in Section III, the packet
delay is given by
\begin{equation}\label{sdelay}
D_s(m)=\Theta\left(\frac{1}{\sqrt{a_s(m)}}\right).
\end{equation}
\end{theorem}

Combining the results in~(\ref{sthroughput1}) and (\ref{sdelay}),
the delay-throughput tradeoff for the secondary tier is given by
the following theorem.
\begin{theorem}
With the secondary protocol defined in Section III, the
delay-throughput tradeoff is
\begin{equation}
D_s(m)=\Theta(m\lambda_s(m)),~\textrm{for}~\lambda_s(m)=O\left(\frac{1}{\sqrt{m\log
m}}\right).
\end{equation}
\end{theorem}
For detailed proofs of the above theorems, please refer
to~\cite{Yin:Scaling}.

\section{Delay and Throughput Analysis for the Primary Tier}

In this section, we first give the throughput and delay scaling
laws for the primary tier, followed by the delay-throughput
tradeoff. Due to the page limit, we only show the main results and
leave all the proofs to the journal version~\cite{Gao:Delay}.
\vspace{-1 mm}
\subsection{Throughput Analysis for the Primary Tier}
In order to obtain the throughput scaling law, we first give the
following lemmas.
\begin{lemma}\label{lemma1}
The numbers of the primary nodes and secondary nodes in each
primary cell are $\Theta(na_p(n))$ and $\Theta(ma_p(n))$ w.h.p.,
respectively.
\end{lemma}
\begin{lemma}\label{lemma2}
If the secondary nodes compete to be the designated relay nodes
for the primary tier by pretending as primary nodes and $a_p(n)=o(1)$, a randomly
selected designated relay node for the primary packet in each primary
cell is a secondary node w.h.p.~(due to the fact that $m>n$ and $n\rightarrow\infty$) such that all the primary packets are actually carried over by the secondary tier w.h.p..
\end{lemma}
\begin{lemma}\label{lemma3}
With the protocols given in Section III, an active
primary cell can support a constant data rate of $K_1$, where
$K_1>0$ independent of $n$ and $m$.
\end{lemma}
\begin{lemma}\label{lemma4}
With the protocols given in Section III, the secondary
tier can deliver the primary packets to the intended primary
destination node at a constant data rate of $K_2$, where $K_2>0$
independent of $n$ and $m$.
\end{lemma}

Based on \emph{Lemmas 1-4}, we have the following theorem.

\begin{theorem}\label{pthroughput}
With the protocols given in Section III, the primary tier can
achieve the following throughput per S-D pair and sum throughput
w.h.p. when $\beta\geq 2$.
\begin{equation}\label{pthroughput1}
    \lambda_p(n)=\Theta\left(\frac{1}{n a_p(n)}\right)
\end{equation}
and
\begin{equation}\label{pthroughput2}
    T_p(n)=\Theta\left(\frac{1}{a_p(n)}\right),
\end{equation}
where $a_p(n)\geq 2\log n/n$ and $a_p(n)=o(1)$.
\end{theorem}

By setting $a_p(n)=2\log n/n$, the primary tier can achieve the
following throughput per S-D pair and sum throughput w.h.p.:
\begin{equation}
    \lambda_p(n)=\Theta\left(\frac{1}{\log n}\right)
\end{equation}
and
\begin{equation}
    T_p(n)=\Theta\left(\frac{n}{\log n}\right).
\end{equation}
\vspace{-4 mm}
\subsection{Delay Analysis of the Primary Tier}

We focus on the delay performance of the primary tier with the aid
of the secondary tier. In the proposed protocols, we know that the
primary tier pours all the primary packets into the secondary tier
w.h.p. based on \emph{Lemma 2}. In order to analyze the delay of
the primary tier, we have to calculate the traveling time for the $N$
segments of a primary packet to reach the corresponding
intermediate destination node within the secondary tier. Since the
S-D paths for the $N$ segments are the same and an active
secondary cell (outside all the preservation regions) transmits
one packet for each S-D path passing through it within a secondary
time slot, we can guarantee that the $N$ segments depart from the
$N$ nodes, move hop by hop along the S-D paths, and finally
reach the corresponding intermediate destination node in a
synchronized fashion. According to the definition of packet
delay, the $N$ segments
experience the same delay given in~(\ref{sdelay}) within the
secondary tier, and all the segments arrive the intermediate destination node within one secondary slot.

Let $L_p$ and $L_s$ denote the durations of the primary and
secondary time slots, respectively. According to the proposed
protocols, we have
\begin{equation}
    L_p=64L_s.
\end{equation}
Since we split the secondary time frame into three fractions and
use one of them for the primary packet transmissions, each primary
packet suffers from the following delay
\begin{equation}\label{pdelay}
    D_p(n)=\frac{3}{64}D_s(m)+C=\Theta\left(\frac{1}{\sqrt{a_s(m)}}\right)
\end{equation}
where $C$ denotes the average time for a primary packet to travel
from the primary source node to the $N$ secondary relay nodes and
from the intermediate destination node to the final destination
node, which is a constant. We see from (\ref{pdelay}) that the
delay of the primary tier is only determined by the size of the
secondary cell $a_s(m)$. In order to obtain a better delay
performance, we should make $a_s(m)$ as large as possible.
However, a larger $a_s(m)$ results in a decreased throughput per
S-D pair in the secondary tier and hence a decreased throughput
for the primary tier since all the primary traffic traverses over
the secondary tier w.h.p.. In the following, we derive the
relationship between $a_p(n)$ and $a_s(m)$ in our supportive
two-tier setup.

We know that given $a_p(n)\geq 2\log n/n$, the maximum throughput
per S-D pair for the primary tier is
$\Theta\left(\frac{1}{na_p(n)}\right)$. Since a primary packet is
divided into $N$ segments and then routed by $N$ parallel S-D
paths within the secondary tier, the supported rate for each
secondary S-D pair is required to be
$\Theta\left(\frac{1}{Nna_p(n)}\right)=\Theta\left(\frac{\sqrt{\log
m}}{\sqrt mna_p(n)}\right)$. As such, based
on~(\ref{sthroughput1}), the corresponding secondary cell size
$a_s(m)$ needs to be set as
\begin{equation}\label{cellsize}
    a_s(m)=\frac{\beta^2 n^2a_p^2(n)}{2 m\log m}
\end{equation}
where we have $a_s(m)\geq 2\log m/m$ when $a_p(n)\geq 2\log n/n$.

Substituting~(\ref{cellsize}) in (\ref{pdelay}), we have the following theorem.
\begin{theorem}
According to the proposed protocols in Section III, the primary
tier can achieve the following delay w.h.p. when $\beta\geq 2$.
\begin{equation}\label{pdelayfinal}
D_p(n)=\Theta\left(\frac{\sqrt{m\log
m}}{na_p(n)}\right)=\Theta\left(\frac{\sqrt{n^\beta\log
n}}{na_p(n)}\right).
\end{equation}
\end{theorem}
\vspace{-1 mm}
\subsection{Delay-throughput Tradeoff for the Primary Tier}
Combining the results in~(\ref{pthroughput1}) and
(\ref{pdelayfinal}), the delay-throughput tradeoff for the primary
tier is given by the following theorem.
\begin{theorem}\label{ptradeoff}
With the protocols given in Section III, the delay-throughput
tradeoff in the primary tier is given by
\begin{equation}\label{tradeoff}
   \hspace{-3 mm} D_p(n)=\Theta\left(\sqrt{n^\beta\log n}\lambda_p(n)\right)~\textrm{for}~\lambda_p(n)=O\left(\frac{1}{\log n}\right).
\end{equation}
\end{theorem}
\vspace{-3 mm}
\section{Conclusion}
In this paper, we studied the throughput scaling laws for an
interactive two-tier network, where the secondary tier is willing
to route packets for the primary tier in an opportunistic fashion.
When the secondary tier has a much higher density, with our
proposed schemes, the primary tier can achieve a better throughput
scaling law of $\Theta\left(1/\log n\right)$ per S-D pair compared
to $\Theta\left(1/\sqrt{n\log n}\right)$ in for non-interactive
overlaid networks with a different delay-throughput tradeoff.

\end{document}